# The Recurrence Relation of Irreducible Tensor Operators for O(4)

Chin-Sheng Wu
Center for General Education
Yuan-Ze University, Taiwan

We derive the recurrence relation of irreducible tensor operator for $O(4)$ in using the Wigner-Eckart theorem. The physical process like radiative transitions in atomic physics, nuclear transitions between excited nuclear states can be described by the matrix element of an irreducible tensor, which is expressible in terms of a sum of products of two factors, one is a symmetry-related geometric factor, the Clebsch-Gordan coefficients, and the other is a physical factor, the reduced matrix elements. The specific properties of the states enter the physical factor only. It is precisely this fact that makes the Wigner-Eckart theorem invaluable in physics. Often time one is interested in ratio of two transition matrix element where it is sufficient to regard only the Clebsch-Gordan coefficients. In this paper we first get the commutation relations of $O(4)$, and then we choose one of these relations to operate on the certain eigenvectors. Finally we take summation over all possible eigenvectors and obtain rather the compact recurrence relation for irreducible tensor operators.

## I. INTRODUCTION

It is well known that group theory has numerous applications in the various fields of theoretical physics[1-2] The degeneracy of the energy levels in a 1/r potential, such as in the hydrogen atom, are related to symmetries of the Schrodinger equation with this potential in a four-dimensional orthogonal space, $O(4)$. The additional one is due to the eccentricity of the orbit, which was discussed the first time by Laplace[3] in 1799.

We derive the recurrence relation of irreducible tensor operator for $O(4)$ in using the Wigner-Eckart theorem. The physical process like radiative transitions in atomic physics, nuclear transitions between excited nuclear states[4] can be described by the matrix element of an irreducible tensor, which is expressible in terms of a sum of products of two factors, one is a symmetry-related geometric factor, the Clebsch-Gordan coefficients, and the other is a physical factor, the reduced matrix elements. The specific properties of the states enter the physical factor only. It is precisely this fact that makes the Wigner-Eckart theorem invaluable in physics. Often time one is interested in ratio of two transition matrix element where it is sufficient to regard only the Clebsch-Gordan coefficients. In this paper we first get the commutation relations of *O*(4), and then we choose one of these relations to operate on the certain eigenvectors. We take summation over all possible eigenvectors. In the end we obtain rather the compact recurrence relation for irreducible tensor operators.

## II. COMMUTATION RELATION

$E_{ij}$ is a matrix with unit in the i$^{th}$ row and j$^{th}$ column, zero for the rest of elements.

$$I_{ij} = E_{ij} - E_{ji}.$$

Chevalley bases[5] of *O*(4), which subject to Cartan-Weyl commutation relations are given as follows:

$$J_0 = iI_{21},$$
$$J_{\pm 1} = -\sqrt{\frac{1}{2}}(I_{31} \pm iI_{32}),$$
$$T_0 = -iI_{43}$$
$$T_{\pm 1} = -\sqrt{\frac{1}{2}}(I_{42} \mp I_{41}),$$

$$[J_0, J_{\pm 1}] = \pm J_{\pm 1},$$
$$[J_{+1}, J_{-1}] = -J_0,$$
$$[T_0, T_{\pm 1}] = \pm J_{\pm 1},$$
$$[T_1, T_{-1}] = -J_0,$$
$$[J_0, T_q] = qT_q,$$
$$[J_{\pm 1}, T_q] = \mp\sqrt{\frac{1}{2}}\sqrt{(1 \mp q)(1 \pm q + 1)}T_{q+1}.$$

## III. CALCULATION AND RESULT

We let

$$[T_1^1, T_{-1}^1] = -J_0,$$

$$T_1^1 T_{-1}^1 - T_{-1}^1 T_1^1 = -J_0$$

operate on $\langle l-1, l-1 |$ and $| l, l-1 \rangle$.

Firstly, we evaluate $\langle l-1, l-1 | T_1^1 | l', m' \rangle \langle l', m' | T_{-1}^1 | l, l-1 \rangle$ by the Wigner-Eckart

theorem.

$$\langle l-1,l-1,1,1|l',m'\rangle\langle l-1\|T\|l'\rangle\langle l,l-1,1,1|l',m'\rangle^*\langle l\|T\|l'\rangle^*,$$

$m_1 = l-1, m_2 = 1, m = m_1 + m_2 = l-1+1 = l, j_1 = l-1.$

If [6] $j = j_1 + 1 = l-1+1 = l$ then

$$\langle l-1,l-1,1,1|l,l\rangle = \sqrt{\frac{(l-1+l)(l-1+l+1)}{(2l-2+1)(2l-2+2)}} = 1.$$

If $j = j_1 = l-1$ then

$$\langle l-1,l-1,1,1|l-1,l\rangle = \sqrt{\frac{(l-1+l)(l-1-l+1)}{(2l-2)(l-1+1)}} = 0.$$

If $j = j_1 - 1 = l-2$ then

$$\langle l-1,l-1,1,1|l-2,l\rangle = \sqrt{\frac{(l-1-l)(l-1-l+1)}{(2l-2)(2l-2+1)}} = 0.$$

$$\langle l,l-1,1,1|l',m'\rangle^*\langle l\|T\|l'\rangle^*,$$

$m_1 = l-1, m_2 = 1, m = m_1 + m_2 = l-1+1 = l, j_1 = l.$

If $j = j_1 + 1 = l+1$ then

$$\langle l,l-1,1,1|l+1,l\rangle = \sqrt{\frac{(l+l)(l+l+1)}{(2l+1)(2l+2)}} = \sqrt{\frac{l}{l+1}}.$$

If $j = j_1 = l$ then

$$\langle l,l-1,1,1|l,l\rangle = -\sqrt{\frac{(l+l)(l-l+1)}{2l(l+1)}} = \sqrt{\frac{1}{l+1}}.$$

If $j = j_1 - 1 = l-1$ then

$$\langle l,l-1,1,1|l-1,l\rangle = \sqrt{\frac{(l-l)(l-l+1)}{2l(2l+1)}} = 0.$$

Secondly, we evaluate $\langle l-1,l-1|T^1_{-1}|l',m'\rangle\langle l',m'|T^1_1|l,l-1\rangle$ by the Wigner-Eckart theorem.

$$\langle l-1,l-1,1,-1|l',m'\rangle\langle l-1|T|l'\rangle\langle l,l-1,1,-1|l',m'\rangle^*\langle l|T|l'\rangle^*,$$

$$m_1 = l-1, m_2 = -1, m = m_1 + m_2 = l-1-1 = l-2, j_1 = l-1.$$

If $j = j_1 + 1 = l-1+1 = l$ then

$$\langle l-1,l-1,1,-1|l,l-2\rangle = \sqrt{\frac{(l-1-l+2)(1+1)}{(2l-2+1)(2l-2+2)}} = \sqrt{\frac{1}{(2l-1)l}}.$$

If $j = j_1 = l-1$ then

$$\langle l-1,l-1,1,-1|l-1,l-2\rangle = \sqrt{\frac{(l-1-l+2)(l-1+l-2+1)}{(2l-2)(l-1+1)}} = \sqrt{\frac{1}{l}}.$$

If $j = j_1 - 1 = l-2$ then

$$\langle l-1,l-1,1,-1|l-2,l-2\rangle = \sqrt{\frac{(l-1+l-2+1)(l-1+l-2)}{(2l-2)(2l-2+1)}} = \sqrt{\frac{2l-3}{2l-1}}.$$

$$\langle l,l-1,1,-1|l',m'\rangle^*\langle l|T|l'\rangle^*,$$

$$m_1 = l-1, m_2 = -1, m = m_1 + m_2 = l-2, j_1 = l.$$

If $j = j_1 + 1 = l+1$ then

$$\langle l,l-1,1,-1|l+1,l-2\rangle = \sqrt{\frac{(l+l-2)(l+l-2+1)}{(2l+1)(2l+2)}} = \sqrt{\frac{(2l-2)(2l-1)}{(2l+1)(2l+2)}}.$$

If $j = j_1 = l$ then

$$\langle l,l-1,1,-1|l,l-2\rangle = \sqrt{\frac{(l-l+2)(l+l-2+1)}{2l(l+1)}} = \sqrt{\frac{2l-1}{l(l+1)}}.$$

If $j = j_1 - 1 = l-1$ then

$$\langle l,l-1,1,-1|l-1,l-2\rangle = \sqrt{\frac{(l+l-2+1)(l+l-2)}{2l(2l+1)}} = \sqrt{\frac{(2l-1)(l-1)}{l(2l+1)}}.$$

Finally $T_1^1 T_{-1}^1 - T_{-1}^1 T_1^1 = -J_0$ can be written as follows.

$$\sum_{l',m'} \langle l-1,l-1|T_1^1|l',m'\rangle\langle l',m'|T_{-1}^1|l,l-1\rangle - \langle l-1,l-1|T_{-1}^1|l',m'\rangle\langle l',m'|T_1^1|l,l-1\rangle =$$

$$\langle l-1,l-1|-J_0|l,l-1\rangle = 0,$$

$$\sqrt{\frac{1}{l+1}}\langle l-1|T|l\rangle\langle l|T|l\rangle - \sqrt{\frac{1}{l^2(l+1)}}\langle l-1|T|l\rangle\langle l|T|l\rangle - \sqrt{\frac{(2l-1)(l-1)}{l^2(2l+1)}}\langle l-1|T|l-1\rangle\langle l|T|l-1\rangle = 0,$$

$$\langle l|T|l\rangle - \sqrt{\frac{1}{l^2}}\langle l|T|l\rangle - \sqrt{\frac{(2l-1)(l-1)(l+1)}{l^2(2l+1)}}\langle l-1|T|l-1\rangle = 0,$$

$$\sqrt{\frac{(l^2-1)}{l^2}}\langle l|T|l\rangle - \sqrt{\frac{(2l-1)(l-1)(l+1)}{l^2(2l+1)}}\langle l-1|T|l-1\rangle = 0,$$

$$\langle l|T|l\rangle - \sqrt{\frac{2l-1}{2l+1}}\langle l-1|T|l-1\rangle = 0.$$

This is one of the recurrence relations of the irreducible tensor operator for $O(4)$. The others can be found by having some commutation relations operator on various eigenvectors.

## IV. CONCLUDING REMARK

It is interesting to find that the recurrence relation of irreducible tensor matrix element is more concise than we expect. The ability to evaluate the matrix element makes many calculations possible such as the decompositions of higher rank groups, the transitions of the nuclear states.